\documentclass[a4paper,11pt]{article}
\usepackage{pos}

\title{Chromatic calorimetry using nanocomposites: status and future challenges}

\author*[a]{Anne-Mazarine Lyon}
\author[a]{Etiennette~Auffray}
\author[b]{Marco~Caporale}
\author[c]{Katerina~Decka}
\author[d]{Michael~Doser}
\author[a]{Jean-Louis~Faure}
\author[e]{Greta~Gajdan}
\author[e]{Yacine~Haddad}
\author[c,f]{Jan~Kral}
\author[e]{Joseph~Mariano}
\author[a]{Loris~Martinazzoli}
\author[g]{Emanuele~Mazzola}
\author[c,f]{Eva~Mihokova}
\author[c]{Solangel~Rojas~Torres}
\author[a,g]{Matteo~Salomoni}
\author[a,c]{Vojtech~Zabloudil}

\affiliation[a]{CERN, Switzerland}
\affiliation[b]{University of Bologna, Italy}
\affiliation[c]{Czech Technical University in Prague, Czech Replublic}
\affiliation[d]{Massachusetts Institute of Technology, United States}
\affiliation[e]{University of Bern, Switzerland}
\affiliation[f]{Institute of Physics of the Czech Academy of Sciences,  Czech Replublic}
\affiliation[g]{Università degli Studi di Milano-Bicocca, Italy}

\emailAdd{anne-mazarine.lyon@cern.ch}

\abstract{
The development of advanced detector technologies will play a key role in future high-energy physics experiments, such as the FCC-ee. In this context, chromatic calorimetry is an emerging concept for electromagnetic calorimeters, with the potential to combine the precise energy measurement of homogeneous calorimeters with longitudinal reconstruction of particle showers. The state of the art in chromatic calorimetry is reviewed, including experimental results from proof-of-concept prototypes and simulation studies investigating quantum-dot-based configurations. Recent progress and current challenges in the development of nanocomposites are also discussed. Finally, future prospects for chromatic calorimetry are outlined.
}

\FullConference{43rd International Conference on High Energy Physics (ICHEP 2026)\\
30 July  to 5 August, 2026\\
Natal, Brazil\\}

\begin{document}
\maketitle

\section{Introduction}

Future experiments, such as the FCC-ee at CERN, will address a broad physics programme, motivating continued development of particle detection technologies. The electromagnetic calorimeter will play a key role in achieving these physics objectives and must provide excellent energy resolution, precise timing, and effective particle identification capabilities in a challenging experimental environment. Longitudinal segmentation would also be highly desirable, as it can provide information on shower development and notably improve particle discrimination. 
Several calorimeter concepts and technologies have been proposed to meet these demanding requirements~\cite{FCCeeCalo,FCCeeDetector}.

Chromatic calorimetry is a novel concept that aims to combine the advantages of homogeneous calorimeters with longitudinal segmentation~\cite{Doser2022}. A chromatic calorimeter (CCAL) consists of successive layers along the calorimeter depth, each associated with a distinct optical emission spectrum, such that the longitudinal development of a particle shower is encoded in the wavelengths of the emitted light. Together with a spectrally sensitive readout, these wavelength components can be distinguished to infer the longitudinal shower profile. The spectral information can be provided either by scintillators 
or by wavelength-shifting materials.

A promising approach to chromatic calorimetry relies on nanocomposites, which consist of quantum dots embedded in a host matrix. Quantum dots are semiconductor nanocrystals that exhibit narrow emission spectra, tunable emission wavelengths, and fast optical emission~\cite{QD2015}. These properties make them particularly attractive for the CCAL: their narrow and tunable emission spectra make it possible to combine multiple layers with distinct, spectrally distinguishable signatures, while fast emission has the potential to provide high timing performance.

This paper reviews the state of the art in chromatic calorimetry and the development of nanocomposites. The experimental results obtained with CCAL prototypes are summarised in Section~\ref{section_experimental}, while the use of quantum-dot-based materials for chromatic reconstruction is motivated by the simulation studies presented in Section~\ref{section_simulation}.  Recent developments in nanocomposites are discussed in Section~\ref{section_nanocomposites}, before the main results and future prospects are summarised in Section~\ref{section_summary}.

\section{Experimental studies}\label{section_experimental}
Previous studies were conducted to demonstrate the  concept of chromatic reconstruction~\cite{Arora2024,Arora2025}. In particular, 
a CCAL prototype was assembled using three scintillating materials with distinct emission spectra: GAGG, EJ-262, and EJ-228, with emission peaks at approximately 540, 481, and 391~nm, respectively~\cite{Arora2025}.  The scintillators were interleaved with transparent lead fluoride ($\mathrm{PbF_2}$) crystals that provide the stopping power required for shower development.
The optical readout consisted of a four-channel multi-anode photomultiplier tube (MaPMT), equipped with optical filters to distinguish the light emitted by the different scintillators. The prototype was tested with electron and positively charged pion beams at the CERN SPS, with energies up to 100~GeV.

\begin{figure}[tb!]
\centering
\raisebox{0.5cm}{\includegraphics[width=0.52\textwidth]{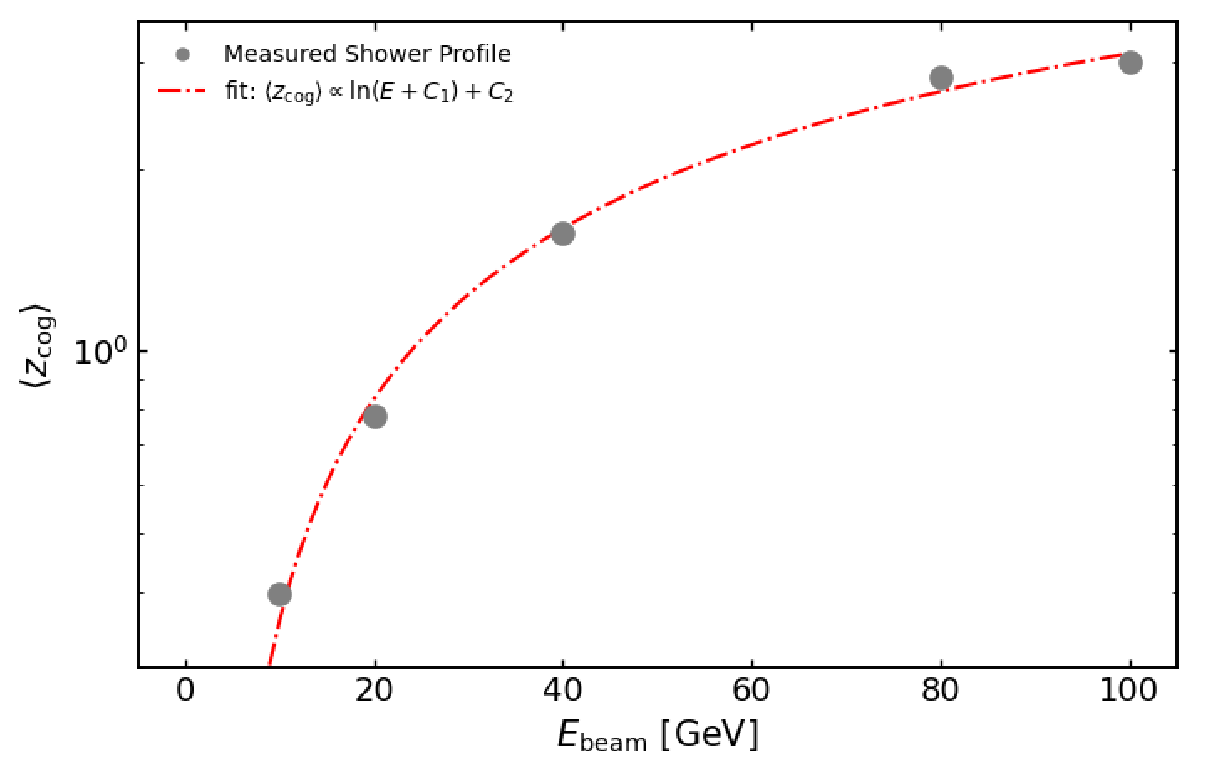}}
\includegraphics[width=0.46\textwidth]{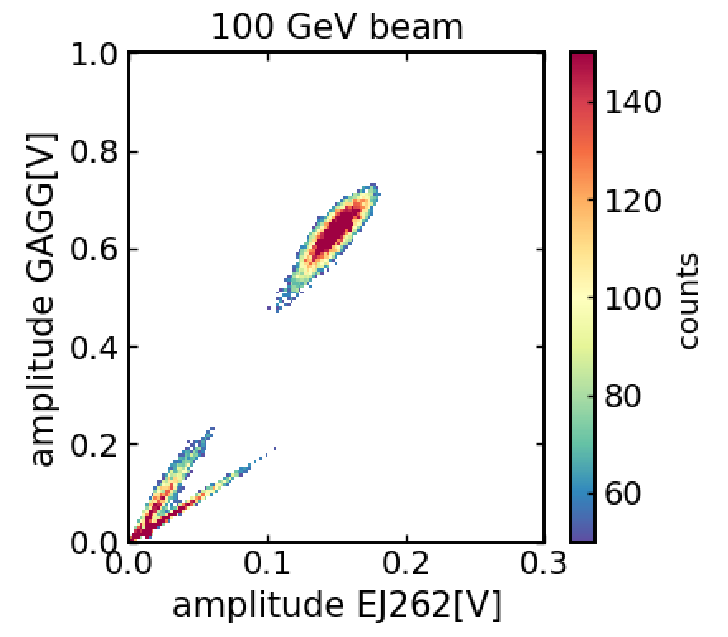}
\caption{\textit{(Left)} Centre of gravity of the electromagnetic shower as a function of the electron beam energy. \textit{(Right)} Signal amplitude in the GAGG layer as a function of that in the EJ-262 layer for electron and charged pion beams at 100~GeV. The clusters in the upper-right and lower-left regions correspond to electrons and charged pions, respectively~\cite{Arora2025}. \label{figure_ccal_2024}}
\end{figure}

Figure~\ref{figure_ccal_2024} summarises the main results. The measured signal amplitudes and their relative contributions across the different readout channels provide information on the longitudinal shower development. In particular, the reconstructed shower centre of gravity exhibits the expected logarithmic dependence on electron beam energy. Furthermore, correlations between the signals from different scintillator layers demonstrate good electron--pion discrimination at 100~GeV. 

Building on the results obtained with this prototype, an upgraded CCAL version
was developed to improve the spectral separation of the scintillation signals and the overall optical and mechanical design. The prototype retains the three scintillators GAGG, EJ-262, and EJ-228, interleaved with transparent PbF$_2$ crystals. 

A major improvement with the upgraded CCAL concerns the optical readout. Unlike the previous prototype~\cite{Arora2025}, which used band-pass filters, 
the upgraded readout employs long-pass filters, which are less sensitive to the angle of incidence. Three filters, with cut-off wavelengths of 320, 475, and 530~nm, respectively, are placed on three of the four MaPMT channels, each collecting a different combination of the scintillator emissions. The individual scintillator contributions are then reconstructed using an optical-response matrix accounting for their emission spectra and light yields, light transport through the stack, MaPMT quantum efficiency, and filter transmittance. This approach improves the robustness of the spectral separation.

Moreover, the upgraded prototype incorporates measures to suppress echo photons, which can arise when light emitted by a downstream scintillator propagates backwards and excites an upstream layer, potentially biasing the reconstructed shower profile. To mitigate this effect, additional long-pass filters are placed along the beam direction before the EJ-262 and EJ-228 layers, with cut-off wavelengths of 530 and 475~nm, respectively.

The optical and mechanical assembly was also improved. The upgraded CCAL prototype is housed in a custom-designed holder wrapped with Enhanced Specular Reflector (ESR), which improves light collection and light tightness. All components are optically coupled using a silicone-based optical coupling compound. The mechanical support ensures reproducible assembly and alignment, facilitating further development towards a scalable detector configuration.

The upgraded prototype was tested with high-energy electron and pion beams at the CERN SPS, with the results to be presented in a separate manuscript. The measurements demonstrate the potential of the improved prototype design for reconstructing longitudinal shower development, particle identification, and linear energy response. These studies represent a further step towards a chromatic calorimeter with improved spectral reconstruction and a more scalable design. 

\section{Simulation studies}\label{section_simulation}

Simulation will play a key role in the transition from CCAL prototypes based on conventional scintillators to configurations using nanocomposites.
It will also be essential for optimising the chromatic stack and readout system. In particular, the choice of materials, the number and thickness of the chromatic layers, and the detector dimensions must be optimised in terms of energy and longitudinal resolution, spectral separation, detection efficiency, and timing performance.

A simulation framework was developed using the Geant4 toolkit~\cite{Geant4}, which enables the simulation of particle interactions and optical photon propagation using the macroscopic properties of the materials. In this framework, the nanocomposites are implemented as wavelength-shifting materials, with their optical behaviour described by parameters including the refractive index, absorption length, and re-emission spectra~\cite{Haddad2025}.

\begin{figure}[tb!]
\centering
\includegraphics[width=0.9\textwidth]{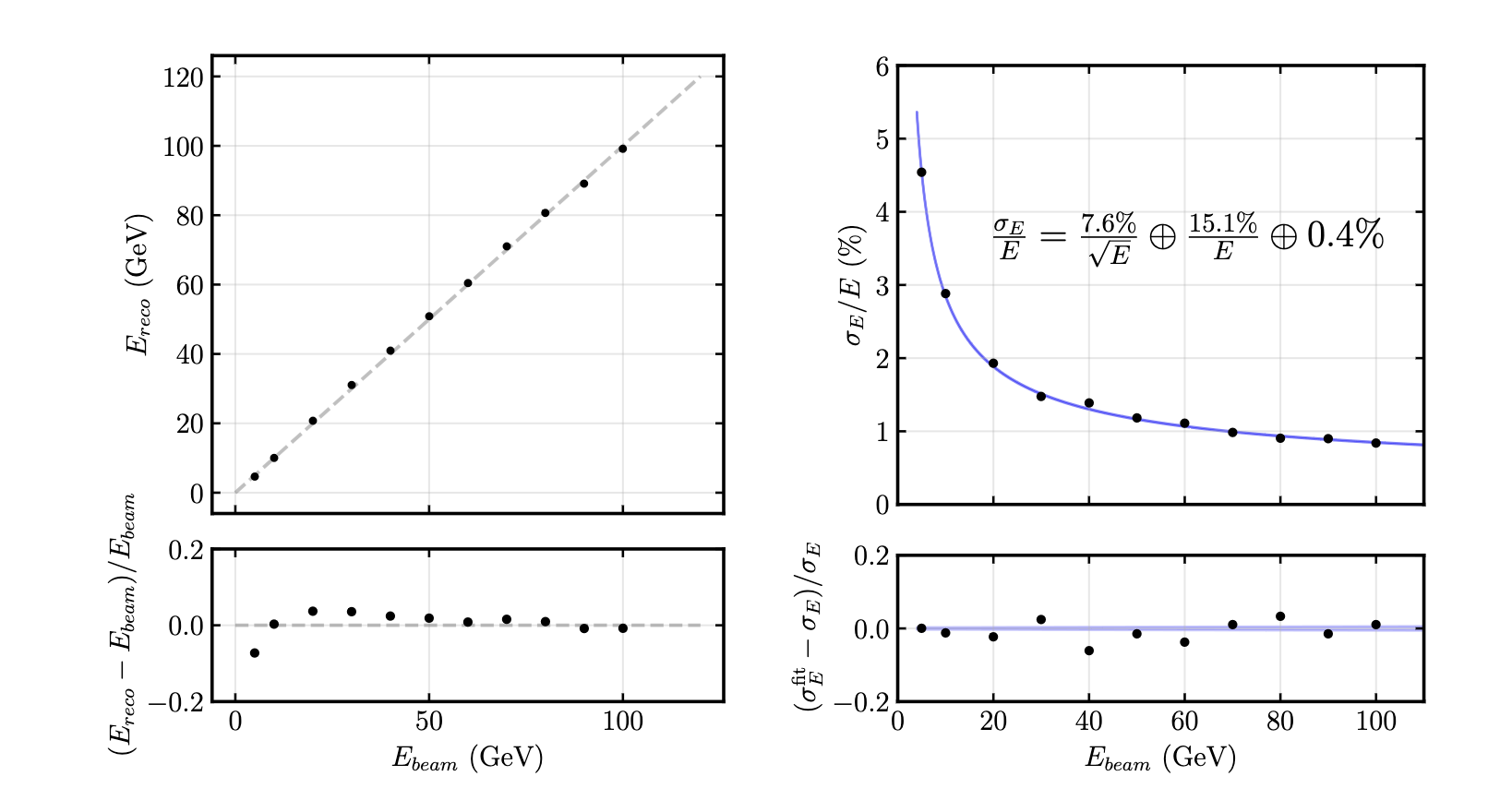}
\caption{Energy response of the simulated quantum-dot-based CCAL prototype, shown in terms of the reconstructed energy \textit{(left)} and energy resolution \textit{(right)} as a function of the incoming beam energy~\cite{Haddad2025}. \label{figure_simulation}}
\end{figure}

The framework was applied to the simulation of a 
CCAL prototype comprising four PbWO$_4$ crystal scintillators interleaved with PMMA-based nanocomposites. Four silicon photomultiplier channels equipped with different optical filters record light in distinct spectral bands, allowing the longitudinal shower profile and energy to be reconstructed from the channel responses. Figure~\ref{figure_simulation} shows that the simulated prototype achieves a linear energy response, with a constant term of approximately 0.4\% in the energy resolution, indicating promising potential for chromatic calorimetry. Although the stochastic term of about 7.6\% is still relatively large, further optimisation of the materials and detector design is expected to improve the overall energy resolution and support further validation against experimental measurements~\cite{Haddad2025}.


\section{Nanocomposite development}\label{section_nanocomposites}
The progress in chromatic calorimetry is closely linked to advances in nanocomposite development. A central challenge is to understand how the energy deposited by an ionising particle is converted into nanocrystal emission. If the host polymer produces prompt Cherenkov light, or if the energy deposited in the host is not efficiently transferred to the quantum dots, the detected signal may contain a substantial contribution unrelated to the desired quantum-dot emission. Carrier trapping
and non-radiative recombination can further reduce the efficiency of this process.

Recent work on cesium lead halide nanocrystals embedded in polymer matrices has demonstrated progress towards high nanocrystal loading, while partially retaining optical transparency~\cite{Kral2025}. Higher loading may increase the probability of energy deposition in the nanocrystals, but it can also enhance self-absorption, scattering, and aggregation.

Nanocomposites prepared using this approach were  tested with high-energy electron beams at the CERN SPS. Preliminary observations indicate that the measured response contains a substantial contribution from Cherenkov light generated in the polymer matrix, while the quantum-dot-related contribution appears comparatively weak. This highlights the need for dedicated measurements to disentangle the different light-production mechanisms and understand the excitation process of nanocrystals under 
particle irradiation.

In view of these challenges, the use of nanocomposites as fast wavelength-shifting layers, rather than bulk primary scintillators, represents a promising direction for chromatic calorimetry.

Ongoing efforts within the DRD5 and Crystal Clear Collaborations at CERN aim to establish a systematic characterisation programme for nanocomposites tailored to high-energy physics applications. This requires optimising the host matrix, quantum-dot composition, surface ligands, and quantum-dot concentration to achieve high light yield and fast timing, while preserving optical transparency and limiting self-absorption.

\section{Summary and outlook}\label{section_summary}

Chromatic calorimetry is an emerging concept for future electromagnetic calorimeters, combining precise energy measurement with longitudinal shower reconstruction and fast timing. Three-layer proof-of-concept prototypes based on commercial crystal and plastic scintillators were assembled and tested to demonstrate the feasibility of chromatic reconstruction. The measurements demonstrated the reconstruction of the shower centre of gravity and the discrimination of electrons from charged pions. The next step is to move beyond the proof of concept towards a full calorimeter module, with a design optimised for spectral separation, energy resolution, and timing performance. 

A Geant4-based simulation framework was also developed to model the detector response and investigate extensions of the CCAL concept. Simulations of a  quantum-dot-based configuration indicate the potential for longitudinal shower reconstruction while offering promising energy resolutions. Further studies will use this framework to optimise material properties, layer configurations, and detector geometry, assessing their impact on the calorimeter performance.

The long-term development of the CCAL will depend on the future advances in nanocomposites, whose tunable emission spectra could notably enable finer spectral separation between longitudinal layers. However, efficient excitation of the quantum dots under ionising radiation, together with the preservation of optical transparency, high light yield, and fast timing, remains challenging. Future material studies will therefore investigate the mechanisms of energy transfer and optical losses, while exploring nanocomposites as fast wavelength-shifting layers. 

Together, these experimental, simulation, and material developments will guide the evolution of chromatic calorimetry from a demonstrated reconstruction principle towards a scalable calorimeter concept for future high-energy physics experiments.

\section*{Acknowledgements}
This work was performed within the framework of the Crystal Clear and DRD5 Collaborations at CERN and the Horizon Europe EIC Pathfinder project UNICORN (GA 101098649), which provided financial support. Additional support was provided by the Grant Agency of the Czech Technical University in Prague (SGS24/145/OHK4/3T/14), the CERN Quantum Technology Initiative, and the Ministry of Education, Youth, and Sports of the Czech Republic (LM2023040). The authors thank the LHCb Picocal team for providing the data acquisition system for the test beams.


\end{document}